\documentclass[a4paper,12pt]{article}
\usepackage{amssymb,amsfonts,amsmath,mathtext,cite,enumerate,float}

\title{On estimating the output entropy of a tensor product of the quantum phase-damping channel with an arbitrary channel}

\author{Grigori G. Amosov\\
{Steklov Mathematical Institute}}

\begin{document}

\maketitle

\begin{abstract}
We obtained the estimation from below for the output entropy of a tensor product of the quantum phase-damping channel with an
arbitrary channel. It is shown that
from this estimation immediately follows that the strong superadditivity of the output entropy holds for this
channel as well as for the quantum depolarizing channel.
\end{abstract}

\section{Introduction}

Let $\mathfrak {S}(H)$ denote the set of all states, i.e. positive unit trace operators, in a Hilbert space $H,\ dimH<+\infty$.
By a quantum channel we mean a completely positive linear map $\Phi :\mathfrak {S}(H)\to \mathfrak {S}(K)$ preserving the trace.
A quantum channel $\Phi $ is said to be unital if $\Phi (\frac {1}{dimH}I_H)=\frac {1}{dimK}I_K$. Here and in the following
we denote $I_L$ the identity operator in a Hilbert space $L$.

Put
$$
S_{min}(\Phi )=\min \limits _{\rho \in \mathfrak {S}(H)}S(\Phi (\rho)),
$$
where $S(\rho)=-Tr(\rho\log\rho)$ is the von Neumann entropy of a state $\rho$.
In \cite{HSH} the quantity
$$
\hat S_{\Phi }(\rho )=\min \limits _{\rho =\sum \limits _i\pi _i\rho _i}\sum \limits _i\pi _iS(\Phi (\rho _i))
$$
was introduced.
We shall say that {\it the strong superadditivity of the output entropy} of the channel $\Phi :\mathfrak {S}(H)\to \mathfrak {S}(H)$ holds
if
\begin{equation}\label{strong}
\hat S_{\Phi \otimes \Omega }(\rho )\ge \hat S_{\Phi }(Tr_K(\rho ))+\hat S_{\Omega }(Tr_H(\rho ))
\end{equation}
for any quantum channel $\Omega :\mathfrak {S}(K)\to \mathfrak {S}(K)$ and states $\rho \in \mathfrak S(H\otimes K)$. In particular, if the strong superadditivity holds for
the channel $\Phi $, then the minimal output entropy is additive with respect to tensor product of channels, i.e.
\begin{equation}\label{add}
S_{min}(\Phi \otimes \Omega)=S_{min}(\Phi)+S_{min}(\Omega)
\end{equation}
is satisfied for all quantum channels $\Omega $. Unfortunately the additivity of minimal output entropy (\ref {add}) is not valid in general \cite {Has}.
Nevertheless, it was proved for many significant cases \cite {King1, King2, Shor, Hol2}.
The strong superadditivity holds for the noiseless channel and for the entanglement-breaking
channels \cite {HSH}.

Notice that the prove of (\ref {add}) in \cite {King1, King2} is based upon the estimation of
the Schatten-von Neumann trace $p$-norms \cite {AHW}. Let us define a quantum relative entropy as follows
$$
S(\rho\ ||\ \sigma)=Tr(\rho \log\rho)-Tr(\rho \log \sigma),
$$
where $\rho ,\sigma \in \mathfrak {S}(H)$. Then, the quantum H-theorem reads \cite {Lin}
\begin{equation}\label{H}
S(\Phi (\rho )\ ||\ \Phi (\sigma ))\le S(\rho\ ||\ \sigma)
\end{equation}
for $\rho ,\sigma \in \mathfrak S(H)$ and for all (not only unital in general) quantum channels $\Phi$.
In \cite {A1, A2} it was introduced the method based upon
the property (\ref {H}). Using this method the additivity in the known cases was proved without estimation of
$p$-norms. The same method allowed to prove the strong superadditivity for the quantum depolarizing channel
\cite {A3}, quantum-classical channels and quantum erasure channels \cite{AM}. Here we will
improve the method introduced in \cite {A1, A2}. It results in the strong estimation from below for the output
entropy of a tensor product of the quantum phase-damping channel with an arbitrary quantum channel.

Pick up an arbitrary orthonormal basis $(e_j)$ in the Hilbert space $H,\ dimH=n<+\infty$.
Suppose that non-negative numbers $\lambda _j\ge 0$ form the probability distribution
on $(e_j)$ such that $\sum \limits _{j=0}^{n-1}\lambda _j=1$. Let us take the discrete Fourier
transforms of $(e_j)$ and $(\lambda _j)$ as follows
$$
\hat \lambda _j=\sum \limits _{m=0}^{n-1}e^{\frac {2\pi i}{n}jm}\lambda _m,
$$
\begin{equation}\label{4}
f_j=\sum \limits _{m=0}^{n-1}e^{\frac {2\pi i}{n}jm}e_m,
\end{equation}
$0\le j\le n-1$. Let us define a linear map $\Psi $ by the formula
$$
\Psi (|f_j><f_k|)=\hat \lambda _{j-k}|f_j><f_k|,
$$
$0\le j,k\le n-1$. The map $\Psi $ determines a quantum channel in $\mathfrak {S}(H)$ for which
only the phases of a state are damped. Thus, $\Psi $ is said to be a {\it phase damping channel}.
Notice that the phase-damping channel introduced in \cite {King2} is a particular case of phase-damping
channels of our definition.

Define the unitary shift operator $V$ as follows
$$
Ve_j = e_{j\ +\ 1\ mod\ n},\ 0\le j\le n.
$$
Then,
\begin{equation}\label{rep}
\Psi (\rho)=\sum \limits _{j=0}^{n-1}\lambda _jV^j\rho V^{*j},
\end{equation}
$\rho \in {\mathfrak S}(H)$.

Given two orthonormal bases $(f_j)$ in $H$ and $(g_j)$ in $K$ the
unit vector $e\in H\otimes K$ can be represented in two different ways, namely
\begin{equation}\label{sch}
|e>=\sum \limits _{j}\mu _j|f_j>\otimes |h_j>
\end{equation}
and
\begin{equation}\label{sch2}
|e>=\sum \limits \nu _j|\tilde h_j>\otimes |g_j>,
\end{equation}
where $\mu _j,\nu _j\in {\mathbb C},\ \sum \limits _j|\mu _j|^2=\sum \limits _j|\nu _j|^2=1$ and
$(h_j),(\tilde h_j)$ are (non-orthogonal in general) unit vectors.
In the following theorem we suppose that $(f_j)$ in (\ref {sch}) are the same as in (\ref {4}). On the other
hand, an orthonormal basis $(g_j)$ in (\ref {sch2}) will be chosen arbitrarily.

{\bf Theorem 1.} {\it For a unit vector $e\in H\otimes K$ represented in (\ref {sch} -- \ref {sch2}) the following
estimation holds
$$
S((\Psi \otimes \Omega )(|e><e|))\ge \sum \limits _j|\nu_j|^2S(\Psi (|\tilde h_j><\tilde h_j|))+\sum \limits _{j}|\mu _j|^2S(\Omega (|h_{j}><h_{j}|),
$$
where $\Omega $ is an arbitrary quantum channel.
Moreover,
$$
Tr_H(|e><e|)=\sum \limits _{j}|\mu_{j}|^2|h_{j}><h_{j}|,
$$
$$
Tr_K(|e><e|)=\sum \limits _j|\nu _j|^2|\tilde h_j><\tilde h_j|.
$$
}

{\bf Corollary 2.} {\it
The strong superadditivity of the output entropy holds for quantum phase-damping channel $\Psi$.
}

{\bf Remark.} {\it The phase-damping channel is known to be complementary to the entanglement-breaking
channel. Thus, the strong superadditivity for this channel follows from \cite {HSH}.}

Now, let us consider the quantum depolarizing channel $\Upsilon $ defined by the formula
$$
\Upsilon (\rho)=(1-p)\rho +\frac {p}{n}I_H,
$$
where $\rho \in \mathfrak {S}(H),\ 0<p\le \frac {n^2}{n^2-1}$.

{\bf Corollary 3.} {\it
The strong superadditivity of the output entropy holds for the quantum depolarizing channel $\Upsilon$.
}

\section {Estimation of output entropy}

To prove Theorem 1 we need the following two Propositions.

{\bf Proposition 4.} {\it Denote $d$ a number of non-zero terms in (\ref {sch}) such that $d\le n$. Then, the
orthogonal projection
$$
P=\sum \limits _{j:\ \mu _j\neq 0}|f_j><f_j|\otimes |h_j><h_j|
$$
on the subspace ${\mathcal L}\subset H\otimes K$ of the dimension $dim{\mathcal L}=d$ has the property
\begin{equation}\label{prop}
(\Psi \otimes Id)(|e><e|)P=P(\Psi \otimes Id)(|e><e|)=(\Psi \otimes Id)(|e><e|),
\end{equation}
i.e. $\mathcal L$ contains the support of the state $(\Psi \otimes Id)(|e><e|)$.
}

Proof of Proposition 4.

It follows from (\ref {sch}) that
$$
P|e><e|=|e><e|P=|e><e|.
$$
On the other hand,
$$
VPV^*=P
$$
because $V|f_j><f_j|V^*=|f_j><f_j|$ for any $j$.
Thus, the result follows from the representation (\ref {rep}). $\Box $

{\bf Proposition 5.} {\it Given $e\in H\otimes K$ represented as (\ref {sch}) the following
estimation holds
$$
S((\Psi \otimes \Omega )(|e><e|))\ge S((\Psi \otimes Id)(|e><e|))+\sum \limits _{j}|\mu _{j}|^2S(\Omega (|h_{j}><h_{j}|),
$$
where $\Omega $ is an arbitrary quantum channel and
$$
Tr_H((|f_{j}><f_{j}|\otimes I_K)|e><e|)=|\mu _j|^2|h_{j}><h_{j}|
$$
such that
$$
Tr_H(|e><e|)=\sum \limits _{j}|\mu _j|^2|h_{j}><h_{j}|
$$
with the vectors $(f_j)$ defined by (\ref {4}).
}

Proof of Proposition 5.

Let us define a state $\rho$ as follows
$$
\rho =(\Psi \otimes Id)(|e><e|)=\sum \limits _{j=0}^{n-1}\lambda _j(V^j\otimes I_K)|e><e|(V^{*j}\otimes I_K),
$$
Put
\begin{equation}\label{proj}
\sigma =\frac {1}{d}P,
\end{equation}
where $d=dimP$ and $P$ was defined in Proposition 4.

Then, the quantum H-theorem implies
\begin{equation}\label{h}
S((Id\otimes \Omega )(\rho )\ ||\ (Id\otimes \Omega )(\sigma))\le S(\rho\ ||\ \sigma ).
\end{equation}
Notice that
$$
S(\rho\ ||\ \sigma )=Tr(\rho \log \rho)-Tr(\rho \log \sigma )=-S((\Psi \otimes Id)(|e><e|))
$$
$$
-Tr(\frac {1}{n}\sum \limits _{j=0}^{n-1}\lambda _j(V^j\otimes I_K)|e><e|(V^{*j}\otimes I_K)
\log \frac {1}{d}P)=
$$
\begin{equation}\label{1}
-S((\Psi \otimes Id)(|e><e|))+\log d
\end{equation}
because (\ref {prop}) is valid in virtue of Proposition 4.

Now we need to calculate $ S((Id\otimes \Omega )(\rho )\ ||\ (Id\otimes \Omega )(\sigma))$.
Taking into account (\ref {prop}) we can conclude
that the state $(\Psi \otimes Id)(|e><e|)$ can be represented as the sum
\begin{equation}\label{rep}
(\Psi \otimes Id)(|e><e|)=\sum \limits _{k,l}|f_{k}><f_{l}|\otimes y_{kl},
\end{equation}
where $y_{kl}$ are operators in $K$ defined by the formula
\begin{equation}\label{y}
y_{kl}=\hat \lambda _{k-l}\mu _k\mu _l|h_k><h_l|
\end{equation}
supported by ${\mathcal L}$ from Proposition 4 and
satisfying the relation
$$
y_{kl}|h_{l}><h_{l}|=|h_{k}><h_{k}|y_{kl}=y_{kl}
$$
and $y_{kl}\neq 0$ only if $\mu _k\neq 0,\mu _l\neq 0$ in (\ref {sch}).
In particular,
\begin{equation}\label{rav}
y_{kk}|h_{k}><h_{k}|=|h_{k}><h_{k}|y_{kk}=y_{kk}.
\end{equation}
Following the definition of $P$ we get
$$
-Tr((\Psi \otimes \Omega )(|e><e|)\log \frac {1}{d}(Id\otimes \Omega )(\sigma ))=
$$
$$
-Tr(\sum \limits _{k,l}|f_{k}><f_{l}|\otimes \Omega (y_{kl}) \log \frac {1}{d}\sum \limits _{k}|f_{k}><f_{k}|\otimes \Omega (|h_{k}><h_{k}|))=
$$
\begin{equation}\label{prom}
-\sum \limits _{k}Tr(\Omega (y_{kk})\log (\Omega (|h_{k}><h_{k}|))+\log d.
\end{equation}
Definition (\ref {y}) implies that
$$
y_{kk}=|\mu _k|^2|h_{k}><h_{k}|
$$
and
$$
|\mu _k|^2=Tr(y_{kk}).
$$
It follows that the terms of a sum in (\ref {prom}) can be rewritten as follows
$$
-Tr((\Psi \otimes \Omega )(|e><e|)\log \frac {1}{d}(Id\otimes \Omega )(\sigma ))=
$$
\begin{equation}\label{2}
-\sum \limits _k|\mu_{k}|^2Tr(\Omega (|h_{k}><h_{k}|)\log (\Omega (|h_{k}><h_{k}|))+\log d.
\end{equation}
Substituting in (\ref {h}) equalities (\ref {1}) and (\ref {2})
we complete the proof.

$\Box $

Proof of Theorem 1.

The noiseless channel $Id$ is a partial case of the phase-damping channel. Hence, applying Proposition 5
to the quantity $S((\Psi \otimes Id)(|e><e|))$ we obtain the result.

$\Box $

Proof of Corollary 2.

Let us consider the quantity
\begin{equation}\label{q}
\hat S_{\Psi \otimes \Omega }(\rho )=\min \limits _{\rho =\sum \limits _s\pi _s\rho _s}\sum \limits _s
\pi _sS((\Psi \otimes \Omega )(\rho _s))
\end{equation}
The minimum in (\ref {q}) is achieved for some set of pure states $\rho _s=|e^{(s)}><e^{(s)}|$. Applying
Theorem 1 to the values $S((\Psi \otimes \Omega )(|e^{(s)}><e^{(s)}|))$
we obtain the following estimation
$$
\hat S_{\Psi \otimes \Omega }(\rho )\ge \sum \limits _{s}\pi _s\sum \limits _j|\nu _{j}^{(s)}|^2S(\Psi (|\tilde h_{j}^{(s)}><\tilde h_{j}^{(s)}|))
$$
$$
+\sum \limits _s\pi _s\sum \limits _{j}|\mu _{j}^{(s)}|^2S(\Omega (|h_{j}^{(s)}><h_{j}^{(s)}|))\equiv C,
$$
where
$$
Tr_H(|e^{(s)}><e^{(s)}|)=\sum \limits _{j}|\mu _{j}^{(s)}|^2|h_{j}^{(s)}><h_{j}^{(s)}|
$$
and
$$
Tr_K(|e^{(s)}><e^{(s)}|)=\sum \limits _j|\nu _j^{(s)}|^2|e_j^{(s)}><e_j^{(s)}|.
$$
Notice that
$$
\sum \limits _s\pi _s\hat S_{\Psi}(Tr_K(|e^{(s)}><e^{(s)}|))\ge \hat S_{\Psi}(Tr_K(\rho )),
$$
by the definition of $\hat H_{\Psi}$,
and
$$
\sum \limits _s\pi _s\sum \limits _{j}|\nu _{j}^{(s)}|^2Tr_H(|h_{j}^{(s)}><h_{j}^{(s)}|)=Tr_H(\rho ).
$$
Then, notice that
$$
\sum \limits _s\pi _s\sum \limits _{j}|\mu _{j}^{(s)}|^2S(\Omega (|h_{j}^{(s)}><h_{j}^{(s)}|))\ge \hat S_{\Omega }(Tr_H(\rho )).
$$
Hence
$$
C\ge \hat S_{\Psi }(Tr_K(\rho))+\hat S_{\Omega }(Tr_H(\rho )).
$$
$\Box $

Proof of Corollary 3.

Following \cite {King2} let us define the set of orthonormal bases $(f_j^{k})$ as follows
$$
|f_j^k>=\sum \limits _{s=0}^{n-1}exp(i\frac {2\pi s^2k}{2d^2})exp(i\frac {2\pi j}{d})|e_s>,
$$
$1\le k\le 2n^2$.
Then, put
$$
U=\sum \limits _{s=0}^{n-1}e^{\frac {2\pi i}{n}s}|e_s><e_s|,
$$
$$
V_k=\sum \limits _{s=0}^{n-1}e^{\frac {2\pi i}{n}s}|f_s^k><f_s^k|,
$$
$1\le k\le 2n^2$.
Now, consider phase-damping channels determined as follows
$$
\Upsilon _k(\rho )=(1-\frac {n-1}{n}p)\rho +\frac {p}{n}\sum \limits _{s=1}^{n-1}V_k^s\rho V_k^{*s},
$$
$1\le k\le 2n^2$.
It is straightforward to check that
\begin{equation}\label {need}
\Upsilon _k(|e_j><e_j|)=|e_{j+1\ mod\ n}><e_{j+1\ mod\ n}|.
\end{equation}

It is known that \cite {King2}
$$
\Upsilon (\rho)=\frac {1-p}{1+(n-1)(1-p)}\frac {1}{2n}\sum \limits _{k=1}^{2n^2}\Upsilon _k(\rho)
$$
\begin{equation}\label{equ}
+\frac {p}{1+(n-1)(1-p)}\frac {1}{2n^3}\sum \limits _{j=1}^{n-1}\sum \limits _{k=1}^{2n^2}U^j\Upsilon _k(\rho)U^{*j}.
\end{equation}

Let us consider the quantity
\begin{equation}\label{q2}
\hat S_{\Upsilon \otimes \Omega }(\rho )=\min \limits _{\rho =\sum \limits _s\pi _s\rho _s}\sum \limits _s
\pi _sS((\Upsilon \otimes \Omega )(\rho _s))
\end{equation}
The minimum in (\ref {q2}) is achieved for some set of pure states $\rho _s=|e^{(s)}><e^{(s)}|$.
Let us consider the quantity $S((\Upsilon \otimes \Omega )(\rho _s))$.
Below we shall apply Theorem 1 to estimate the quantity $S((\Upsilon \otimes \Omega )(\rho _s))$.
There are two possible decompositions (\ref {sch}) and (\ref {sch2}) of $e^{(s)}$. The decomposition (\ref {sch})
is fixed, while (\ref {sch2}) is depending on a choice of the orthogonal basis $(g_j)$.
We will take $(g_j)$ in such a way that the unit vectors $(\tilde h_j)$ in (\ref {sch2}) would be
orthogonal (it is not so in general). For these purposes
it is appropriate to take $(g_j)$ from the Schmidt decomposition of $|e^{(s)}><e^{(s)}|$ defined by
the formula
\begin{equation}\label{d}
|e^{(s)}>=\sum \limits _j\nu _j^s|\tilde g_j^s>\otimes |g_j^s>,
\end{equation}
$\nu _j^s\ge 0,\ \sum \limits _j|\nu _j^s|^2=1$. Using the covariance property
$$
W\Upsilon (\rho)W^*=\Upsilon (W\rho W^*)
$$
for any $\rho \in \mathfrak {S}(H)$ and any unitary operator $W$ in $H$, we can change
the orthogonal vectors $(\tilde g_j^s)$ in (\ref {d}) to the vectors $(e_j)$ satisfying
the property (\ref {need}). In fact, it suffices to put $W\tilde g_j^s=e_j$.
Thus, we obtain
\begin{equation}\label{equa}
S((\Upsilon \otimes \Omega )(\rho _s))=S((\Upsilon \otimes \Omega )(|\tilde e^{(s)}><\tilde e^{(s)}|)),
\end{equation}
where
$$
|\tilde e^{(s)}>=\sum \limits _j\nu _j^s|e_j>\otimes |g_j^s>
$$
like in (\ref {sch2}) with $\tilde h_j=e_j$.
Using the representation (\ref {equ}) by concavity of entropy we conclude
$$
S((\Upsilon \otimes \Omega )(\rho _s))\ge \frac {1-p}{1+(n-1)(1-p)}\frac {1}{2n}\sum \limits _{k=1}^{2n^2}S((\Upsilon _k\otimes \Omega )(|\tilde e^{(s)}><\tilde e^{(s)}|))+
$$
\begin{equation}\label{d3}
\frac {p}{1+(n-1)(1-p)}\frac {1}{2n^3}\sum \limits _{j=1}^{n-1}\sum \limits _{k=1}^{2n^2}S((\Upsilon _k\otimes \Omega )(|\tilde e^{(s)}><\tilde e^{(s)}|)).
\end{equation}
Applying Theorem 1 to the quantities $S((\Upsilon _k\otimes \Omega )(|\tilde e^{(s)}><\tilde e^{(s)}|))$ we get
\begin{equation}\label{d1}
S((\Upsilon _k\otimes \Omega )(\rho _s))\ge \sum \limits _j|\nu _j^s|^2S(\Upsilon _k(|e_j><e_j|))
\end{equation}
$$
+\sum \limits _j|\mu _j^s|^2S(\Omega (|h_j^s><h_j^s|)),
$$
while
\begin{equation}\label{d2}
S(\Upsilon _k(|e_j><e_j|))=-(1-\frac {n-1}{n}p)\log (1-\frac {n-1}{n}p)-
\frac {p(n-1)}{n}\log \frac {p}{n}
\end{equation}
for all $k$ due to the property (\ref{need}) and
\begin{equation}\label{eee}
Tr_H(|e^{(s)}><e^{(s)}|)=Tr_H(\rho _s)=\sum \limits _j|\mu _j^s|^2|h_j^s><h_j^s|,
\end{equation}
where $(h_j^s)$ and $(\mu _j^s)$ are taken from the representation (\ref {sch}) of $\tilde e^{(s)}$.
Substituting (\ref {d2}) to (\ref {d1}) and (\ref {d1}) to (\ref {d3})
we obtain
\begin{equation}\label{en}
S((\Upsilon \otimes \Omega )(\rho _s))\ge -(1-\frac {n-1}{n}p)\log (1-\frac {n-1}{n}p)-
\frac {p(n-1)}{n}\log \frac {p}{n}+
\end{equation}
$$
+\sum \limits _j|\mu _j^s|^2S(\Omega (|h_j^s><h_j^s|)).
$$
Notice that
$$
-(1-\frac {n-1}{n}p)\log (1-\frac {n-1}{n}p)-
\frac {p(n-1)}{n}\log \frac {p}{n}=
$$
$$
S(\Upsilon (|g><g|))=const
$$
for any unit vector $g\in H$. It follows that
\begin{equation}\label {zan1}
\hat S_{\Upsilon }(\rho )=-(1-\frac {n-1}{n}p)\log (1-\frac {n-1}{n}p)-
\frac {p(n-1)}{n}\log \frac {p}{n}=const
\end{equation}
for all states $\rho \in \mathfrak {S}(H)$.
On the other hand,
\begin{equation}\label{zan2}
\sum \limits _j|\mu _j^s|^2S(\Omega (|h_j^s><h_j^s|))\ge \hat S_{\Omega }(Tr_H(|e^{(s)}><e^{(s)}|))
\end{equation}
due to (\ref {eee}).
Taking into account (\ref {zan1}) and (\ref {zan2}) we conclude that (\ref {en}) gives rise to the inequality
$$
S((\Upsilon \otimes \Omega )(\rho _s))\ge \hat S_{\Upsilon}(Tr_K(\rho _s))+\hat S_{\Omega }(Tr_H(\rho _s)).
$$
Now, to complete the proof it suffices to notice that the minimum in (\ref {q2}) is achieved for the set of states
$\rho _s=|e^{(s)}><e^{(s)}|$.
$\Box $

\section*{Acknowledgments} The author is grateful to A.S. Holevo and all participants of the seminar
"Quantum Probability, Statistics and Information" in Steklov Mathematical Institute for fruitful discussions.
The work is partially supported by Fundamental Research Program of RAS and the RFBR grants 12-01-00319-à and 11-02-00456-à.


\begin{thebibliography}{99}

\bibitem{AHW} G.G. Amosov, A.S. Holevo, R.F. Werner, "On some
additivity problems in quantum information theory",
Probl. Inf. Transm. 36 (2000) 24-34.

\bibitem{A1} G.G. Amosov, "Remark on the additivity conjecture for
the depolarizing quantum channel", Probl. Inf. Transm.
42 (2006) 3-11.

\bibitem{A2} G.G. Amosov, "On the Weyl channels being covariant with
respect to the maximum commutative group of unitaries",
J. Math. Phys. 48 (2007) 2104-2117.

\bibitem{A3} G.G. Amosov, "The strong superadditivity conjecture holds for the quantum depolarizing channel in
any dimension",  Physical Review A 75 (2007) no. 6, P. 060304.

\bibitem{AM} G.G. Amosov, S. Mancini, "The decreasing property of relative entropy
and the strong superadditivity of quantum channels", Quantum Information and Computation 7 (2009) 594-609.


\bibitem{Has} M. B. Hastings, "A Counterexample to Additivity of Minimum Output Entropy", Nature
Physics 5, 255 - 257 (2009), arXiv:0809.3972v3

\bibitem{Hol2} A. S. Holevo, "On complementary channels and the additivity problem", Probab. Theory and
Appl., 51, 133-143, (2005).

\bibitem{HSH} A.S. Holevo, M.E. Shirokov, "On Shor's channel extension
and constrained channels", Commun. Math. Phys.
249 (2004) 417-436.

\bibitem{King1} C. King, "Additivity for unital qubit channels", J. Math. Phys. 43 4641-4653 (2002).

\bibitem{King2} C. King, "The capacity of the quantum depolarizing channel", IEEE Trans. Info. Theory 49,
221-229 (2003).


\bibitem{Lin} G. Lindblad, "Completely positive maps and entropy inequalities"
Commun. Math. Phys. 40 (1975) 147-151.

\bibitem{Shor} P. Shor, "Additivity of the Classical Capacity of Entanglement-Breaking Quantum Channels",
J. Math. Phys. Vol. 43, 4334-4340 (2002)










\end{thebibliography}
\end{document}